\documentclass[lettersize,journal]{IEEEtran}

\usepackage{amsmath,amsfonts,amssymb}
\usepackage{algorithmic}
\usepackage{algorithm}
\usepackage{array}
\usepackage[caption=false,font=footnotesize]{subfig}

\usepackage{textcomp}
\usepackage{stfloats}
\usepackage{url}
\usepackage{verbatim}
\usepackage{graphicx}
\usepackage{cite}
\usepackage{adjustbox}
\usepackage{comment}
\DeclareMathOperator*{\argmax}{arg\,max}

\begin{document}
\title{A Novel Exploitative and Explorative GWO-SVM Algorithm for Smart Emotion Recognition}

% \title{A Novel Exploitative and Explorative GWO-SVM Algorithm for Single Channel ECG based Emotion Recognition}

\author{Xucun Yan, 
        Zihuai Lin,~\IEEEmembership{Senior Member, IEEE,}
        Zhiyun Lin, 
        and~Branka Vucetic,~\IEEEmembership{Life Fellow,~IEEE}
\thanks{Xucun Yan, Zihuai Lin and Branka Vucetic are with School of Electrical and Information Engineering, University of Sydney, New South Wales 2006, Australia (e-mail: xucun.yan@sydney.edu.au; zihuai.lin@sydney.edu.au; branka.vucetic@sydney.edu.au).}
\thanks{Zhiyun Lin is with the Department of Electrical and Electronic Engineering, Southern University of Science and Technology, Shenzhen 518055, China, and Peng Cheng Laboratory, Shenzhen 518066, China (email: linzy@sustech.edu.cn). Corresponding author: Zhiyun Lin}
}

% The paper headers

% \markboth{IEEE Internet of Things Journal,~Vol.~XX, No.~XX, XXXX 2022}%
% {Xucun Yan \MakeLowercase{\textit{et al.}}: A Novel Exploitative and Explorative GWO-SVM Algorithm for Single Channel ECG-based Emotion Recognition}

\maketitle
\begin{abstract}
Emotion recognition or detection is broadly utilized in patient-doctor interactions for diseases such as schizophrenia and autism and the most typical techniques are speech detection and facial recognition. However, features extracted from these behavior-based emotion recognitions are not reliable since humans can disguise their emotions. Recording voices or tracking facial expressions for a long term is also not efficient. Therefore, our aim is to find a reliable and efficient emotion recognition scheme, which can be used for non-behavior-based emotion recognition in real-time. This can be solved by implementing a single-channel electrocardiogram (ECG) based emotion recognition scheme in a lightweight embedded system. However, existing schemes have relatively low accuracy. For instance, the accuracy is about 82.78\% by using a least squares support vector machine (SVM). Therefore, we propose a reliable and efficient emotion recognition scheme---exploitative and explorative grey wolf optimizer based SVM (X-GWO-SVM) for ECG-based emotion recognition. Two datasets, one raw self-collected iRealcare dataset, and the widely-used benchmark WESAD dataset are used in the X-GWO-SVM algorithm for emotion recognition. Leave-single-subject-out cross-validation yields a mean accuracy of 93.37\% for the iRealcare dataset and a mean accuracy of 95.93\% for the WESAD dataset. This work demonstrates that the X-GWO-SVM algorithm can be used for emotion recognition and the algorithm exhibits superior performance in reliability compared to the use of other supervised machine learning methods in earlier works. It can be implemented in a lightweight embedded system, which is much more efficient than existing solutions based on deep neural networks. 

\end{abstract}


\section{Introduction}
\label{intro}

\IEEEPARstart{T}{he} use of the Internet of Things (IoT) is growing steadily over the years. It is expected that by 2025, there will be approximately 27 billion connected IoT devices \cite{IoTanalysis}. At present, the IoT is one of the main promoters of technological innovation and one of the areas with greater potential for social and economic transformation  \cite{leng2020implementation,IoT_FD,RF_energy1}. 
Through a network of sensors and actuators connected to a wireless network \cite{NC1,NC2,NC3,NC4,WRN,distributedRaptor,Raptor_ML,JNCC,RCRC,NC_book}, the operator has the power to remotely gather data. Alternatively, actuators could be programmed to actuate automatically according to values reported by the sensor.

Emotion recognition or detection based on IoT wireless sensing and networking has gained lots of attention since it can be broadly utilized in interfaces between humans and computers and
patient-doctor interactions for diseases such as schizophrenia
and autism. Most emotion detection methods are based on behaviors such as speech detection and face recognition~\cite{speech,d(facial)}. However, features extracted from the abovementioned behavior-based emotion recognition are not adequate for identifying emotions, because the behavior induced by emotion can be disguised by artifacts of human social masking~\cite{hsu2017automatic}. For example, emotion recognition based on facial expressions can be easily misled by a poker face. Using physiological signals, such as electroencephalograms (EEGs)~\cite{EEGxucun,9429732,9793561}, electromyograms (EMGs), and electrocardiograms (ECGs)~\cite{zahid2021robust}, is an alternative to identify emotions since physiological signals are one of the most notable means to manifest the central nervous system in which emotions are processed~\cite{cacioppo2007handbook}. 

Using physiological cues for emotion identification has two advantages over prior approaches to emotion recognition. The first is that physiological signals generated from automatic reactions are difficult to disguise. The second is that wearable emotion monitoring can continually record physiological information. This differs from the instance of voice recognition where data may only be recorded when individuals are speaking. However, using multi-channel biosignals to recognize human emotions is not suitable for practical applications because subjects may be hindered during daily life activities~\cite{5999653}. It has been proved that ECG signals are a suitable physiological channel with acceptable recognition abilities~\cite{hsu2017automatic}.

However, according to~\cite{compare_face_and_ECG,face,ko2018brief,shu2018review}, the accuracy of emotion detection based on a single ECG channel fluctuates a lot for various datasets compared to that of other approaches such as facial emotion recognition. 
On the one hand, recent efforts in emotion recognition using ECG
signals have largely relied on relatively simple supervised learning
techniques~\cite{s21155015}, such as random forest (RF), support vector machine
(SVM), K-nearest neighbor (K-NN), etc. However, these methods
have relatively low accuracy (for instance, the accuracy is about 82.78\% ~\cite{hsu2017automatic} by using least squares SVM).
On the other hand, the current maximum level of single ECG channel-based emotion recognition accuracy reaches 96.9\%~\cite{sarkar2020self} for Wearable Stress and Affect Detection (WESAD) database and 88.2\%~\cite{sarkar2020self} for a dataset for multi-modal research of affect, personality traits, and mood in individuals and groups (AMIGOS)~\cite{shu2018review}, which utilizes self-supervised convolutional neural network (CNN) model. Facial emotion recognition accuracy achieves 92.07\%~\cite{JAIN2018101} for MMI Facial Expression Database and 94.91\%~\cite{JAIN2018101} for the Japanese Female Facial Expression Database, which uses CNN embedded with recurrent neural network (RNN)~\cite{ko2018brief}. 
Nevertheless, these deep neural network-based techniques, e.g., CNN, RNN, etc, tend to
achieve high accuracy but are complex with low computation efficiency, which cannot be implemented in a lightweight embedded system operating in real-time. Therefore, seeking a simple supervised learning scheme to accurately, stably, and efficiently recognize emotions based on a single ECG channel in a lightweight embedded system is necessary.

Towards this objective, this paper aims to develop a novel exploitative and explorative GWO-SVM (X-GWO-SVM) for ECG-based emotion recognition. The goal is to achieve good classification accuracy (as high as utilizing complex neural networks) while simultaneously reducing computation so that it can be implemented in a lightweight embedded system. The idea is motivated from the fact that the SVM algorithm can be used to solve single-channel ECG-based emotion recognition issue with lightweight embedded system implementation. However, the existing SVM works do not offer a good classification accuracy performance for ECG-based recognition due to difficulties in finding appropriate hyper-parameters while preventing overfitting of the training data. 

In general, the selection of hyperparameters is a non-convex optimization issue. Therefore, many heuristic algorithms such as genetic algorithm (GA), particle swarm optimization (PSO), and grey wolf
optimizer (GWO)~\cite{Wainer2021,zhou2021performance,li2020application,wei2017improved} are introduced to tackle it. Compared with PSO and a set of search algorithms, GWO provides better performance in computation reduction (e.g., in feature subset selection~\cite{Emary2015}). Moreover,
the GWO approach has been demonstrated to be more stable against
initialization than PSO and GA~\cite{Emary2015}. However, as discussed in~\cite{zhou2018color},
conventional GWO-based SVM (GWO-SVM) techniques are still
easy to fall into local solutions. 

In this work, an improved method, the X-GWO-SVM method, is proposed. The proposed X-GWO-SVM method is the first to apply GWO-SVM idea to solve ECG-based recognition, and as shown in this paper, this method has higher recognition accuracy than existing SVM and PSO-SVM techniques for ECG emotion recognition use. It can effectively avoid the algorithm falling into a local solution by increasing the exploration ability, and speed up the convergence by increasing the exploitation ability. In this paper, two datasets, one raw self-collected iRealcare dataset, and the widely used benchmark WESAD dataset are used in the X-GWO-SVM algorithm for emotion recognition. Leave-single-subject-out cross-validation yields a mean accuracy of 93.37\% and an F1-score of
93.38\% for the iRealcare dataset and a mean accuracy of 95.93\%
and an F1-score of 95.56\% for the WESAD dataset.

The main contributions of this paper are summarized as follows:
\begin{enumerate}
\item We use a self-built wearable IoT ECG patch with only one ECG channel to collect four emotions, i.e., happiness, tension, peacefulness and excitement, by playing different videos.  

\item  We designed a novel X-GWO-SVM algorithm to internally learn hyperparameters on SVM. It can effectively avoid the algorithm falling into a local solution by increasing the exploration ability, and speed up the convergence by increasing the exploitation ability. 

\item This novel X-GWO-SVM algorithm can accurately and efficiently recognize emotions for single-channel ECG-based signals and be implemented in a lightweight embedded system operating in real-time. It improves accuracy compared to existing simple machine learning methods and dramatically reduces complexity compared to some novel deep neural networks. Thus, the efficiency is also increased compared to other time-consuming emotion recognition methods. 
\end{enumerate}

The outline of the rest of the paper is given below. Section~\ref{Dataset} introduces our database and an expanded dataset. Our model formulation is described in Section~\ref{method}. In Sections~\ref{Classification results} and~Section\ref{discussion}, we present results and discussions, respectively, before concluding with a discussion of potential future directions in Section~\ref{conclusion}.

\section{Dataset}
\label{Dataset}
ECG signals are composed of the P wave, T wave, and QRS complex, which represent the three phases of an ECG pulse. In atrial systole, the P wave is the contraction pulse. The QRS complex signifies ventricular depolarization. The T wave represents ventricular re-polarization~\cite{ECG_characteristics}. An ECG device records the electrical changes caused by the activities of the heart, which are collected by electrodes over the skin for a period of time. It has been proved that ECG signals are a suitable physiological channel with acceptable recognition abilities~\cite{hsu2017automatic} to identify emotions. Therefore, single-channel ECG signals are used in this study. In order to verify the general representation ability of X-GWO-SVM, two datasets of ECG signals are used, one raw self-collected iRealcare dataset with 5 subjects and the other widely-used benchmark WESAD dataset with 15 subjects.

\subsection{Description of iRealcare dataset\label{sec:data_A}}
Data collection is one of the most important steps for emotion detection.
The definition of different emotions must be explicit in this phase.
If the definition is not clear, confusion may occur among different
emotions in the classification phase and the classification performance will be influenced negatively. However, emotions normally instantaneously occur and hold for a short period. The longer the period is, the more irrelevant data is included in ECG signals. Thus, it is hard to properly label the corresponding emotion class.

To avoid the aforementioned issue, we self-collect a dataset with high quality and a short period for each emotion, making sure accurate data collection and labeling processes. The ECG signals are recorded by a low-cost wearable IoT ECG patch, called iRealcare~\cite{IREALCARE1, IREALCARE2, IREALCARE3,IREALCARE4,IREALCARE5} with 128 Hz sampling rates. The data collected by the iRealcare IoT ECG sensor can be transmitted to a smartphone application (APP) via Bluetooth Low Energy (BLE) and then to a cloud. From the cloud, we can acquire the raw ECG signals. Signals are recorded for four emotions including happiness, tension, peacefulness, and excitement.
Except for peacefulness, each emotion is generated based on an external environmental stimulus, which is similar to the published datasets stimulating subjects through audio or video~\cite{schmidt2018introducing}. The peacefulness describes the normal state, for which the ECG signals are recorded without any external stimulus. Signals for happiness, tension, and excitement are recorded when subjects watch comedies, watch thriller movies and do exercises, respectively. Generally, the record duration should be short as we discussed before. Therefore, the record time is in a range of 3.22-6.16 minutes for each emotion. 

It should be noticed that we only record the period that subjects are actually in that emotion
condition and ignore the transition period. Clearly, the definition of different emotions under this external stimulus setting is clear and subjects are easy to get into a specific emotion. Taking into account differences among different subjects, 5 subjects are involved and each subject is recorded with four emotion types. 
For each subject, there are 192-229 samples for peacefulness, 99-141 samples for excitement, 156-236 samples for happiness and 166-205 samples for tension.
More information on the iRealcare database is shown in Table~\ref{Data information of dataset A} and segmentation  details are described in Section~\ref{segmentation and splitting}.

\begin{table*}[h]
\caption{Data information of iRealcare database. }
\begin{adjustbox}{width=6in,center}
\centering
\begin{tabular}{cccccc}
\hline
\hline	
ID	&	Peacefulness duration/min				&	Excitement duration/min				&	Happiness duration/min				&	Tension duration/min				&	Total duration/min				\\
	&	 (Segment number)				&	 (Segment number)				&	 (Segment number)				&	 (Segment number)				&	 (Segment number)				\\
\hline																										
1	&	5	(	192	)	&	3.22	(	123	)	&	6.16	(	236	)	&	4.33	(	166	)	&	18.71	(	717	)	\\
2	&	5.23	(	200	)	&	3.63	(	139	)	&	5.42	(	208	)	&	4.48	(	172	)	&	18.76	(	719	)	\\
3	&	5.37	(	206	)	&	3.69	(	141	)	&	4.91	(	188	)	&	5.35	(	205	)	&	19.32	(	740	)	\\
4	&	5.98	(	229	)	&	3.65	(	140	)	&	4.51	(	173	)	&	4.7	(	180	)	&	18.84	(	722	)	\\
5	&	5.06	(	194	)	&	2.59	(	99	)	&	4.07	(	156	)	&	4.66	(	178	)	&	16.39	(	627	)	\\
\hline																										
Sum	&	26.63	(	1021	)	&	16.78	(	642	)	&	25.08	(	961	)	&	23.52	(	901	)	&	92.01	(	3525	)	\\

\hline	
\hline
\end{tabular}
\label{Data information of dataset A}
\end{adjustbox}
\end{table*}

\subsection{Description of WESAD dataset\label{sec:data_B}}
The dataset, accessible in~\cite{schmidt2018introducing}, is comprised of recordings of 15 subjects (aged 24–35) watching video clips and doing public speaking and mental arithmetic tasks.
 The dataset is recorded with a wrist-based device (including the following sensors: photoplethysmography, accelerometer, electrodermal activity, and body temperature) and a chest-based device (including the following sensors: ECG, accelerometer, electromyogram, respiration, and body temperature). This dataset offers a fusion of physiological parameters to efficiently identify human emotions, as these represent the body's instinctive reactions. However, it is not suitable for practical applications, and it may hinder subjects during daily life activities~\cite{5999653}. Therefore, in this paper, we only study single ECG channel signals for this dataset. The ECG signal is acquired from a RespiBAN Professional using a three-lead configuration with 700$Hz$ sampling rates. Three types of emotions (baseline, stress and amusement) are annotated by subjects~\cite{schmidt2018introducing}. Amusement condition signals are collected when subjects watch funny video clips. Stress condition signals are collected when subjects are asked to provide public speaking. Baseline
condition signals are collected when subjects sit/stand at a table and read magazines. For each subject, there are 9 samples for amusement, 15-18 samples for stress, and 28-29 samples for baseline. The segmentation details are described in Section~\ref{segmentation and splitting}.

\section{Method}
\label{method}
\subsection{Preprocessing}

Normally, ECG signals are non-linear with low signal amplitudes. The frequency range of ECG signals is from 0.05$Hz$ to 100$Hz$ and the dynamic range is below 4$mV$~\cite{all_noise}. Thus, the collected ECG signals are susceptible to being disturbed by external factors such as interference. To acquire ECG signals with low interference, we conduct the pre-processing of the raw ECG signals. During the data collection and transmission stage, ECG signals are mainly affected by baseline drift, power line interference, and electrode contact noise. 
The baseline drift is caused by body movement and breathing. It can make the entire ECG signal shift down or up at the horizontal axis. The frequency of baseline drift is around 0.5$Hz$ and it will influence the analysis of ECG signals.
The power-line interference is characterized by 50 or 60$Hz$, which can be caused by the electromagnetic field of nearby facilities and electromagnetic interference of the power lines. Since the iRealcare sensor used BLE instead of cables, the power-line interference will not affect the collected ECG signals from the sensor. 
The electrode contact noise is caused by the variance of impedance when the skin is stretched. This frequency is typically between 1 and 10$Hz$~\cite{noise}. 

    \subsubsection{Filtering}  
    The finite impulse response (FIR) filter is used to filter the aforementioned noises. It is a reliable and simple filter. Moreover, the output of a FIR filter is not distorted because it is a linear filter~\cite{FIR}. FIR filters are created utilizing window-based techniques, such as the Hamming window, Rectangular window, Hanning window, and the Blackman window. These different windows are used to design the low pass filter and high pass filter with cut-off frequencies. For our band-pass FIR filter, the cut-off frequencies are set to 3$Hz$ and 100$Hz$, respectively. 
    
    \subsubsection{Segmentation and splitting}  
    \label{segmentation and splitting}
    For the iRealcare dataset, the aforementioned 20 groups are denoised, non-overlapping segmented with 200 data points (1.56s), and then split into training and test sets. Non-overlapping is designated between segments to avoid any potential data leakage between training and test data. It should be noticed that the selection of the window size (200 data points) is empirical. 
    Prior research employing these datasets utilized a broad variety of window sizes. For instance,~\cite{schmidt2018introducing} has chosen 5-second windows for WESAD whereas~\cite{lin2019explainable} has used 1-second windows for the same dataset. Specifically, the training set consists of 16 groups, each of which has four emotions, whereas the test set consists of 4 groups. Similar to the iRealcare dataset, the WESAD dataset is also filtered by a FIR filter, non-overlapping segmented with 14000 data points (20s), and then 12 subjects are treated as a training set while the rest 3 subjects form a test set.

    Fig.~\ref{f1} depicts four emotion segments with 200 randomly chosen ECG signal data samples from the iRealcare dataset.
    We can see that for the emotion of peacefulness, the subject's heart rate is comparatively sluggish. However, it is hard to identify the other three emotions based on the original ECG signals. As a result, the design of an efficient feature extraction approach is necessitated.

\subsubsection{Discrete cosine transform (DCT)}
\label{Discrete cosine transform}
In this paper, we use the DCT  methods to extract the main information of ECG signals in the frequency domain~\cite{DCT}. It is computed for a compressed version of input ECG signals containing significant information, and only a small subset of the coefficients is maintained as a feature vector. The main merit of the DCT is its high computational speed which is suitable for data compression~\cite{Hafed2001}.
To improve performance, the $Z$-score normalization technique is invoked prior to recognition to account for small perturbations in motion artifacts caused by electrodes’ movement on the skin surface.

\begin{figure}[htbp]
\centering
\includegraphics[width=3.5in ,trim={3cm 6.5cm 4.2cm 6.5cm},clip]{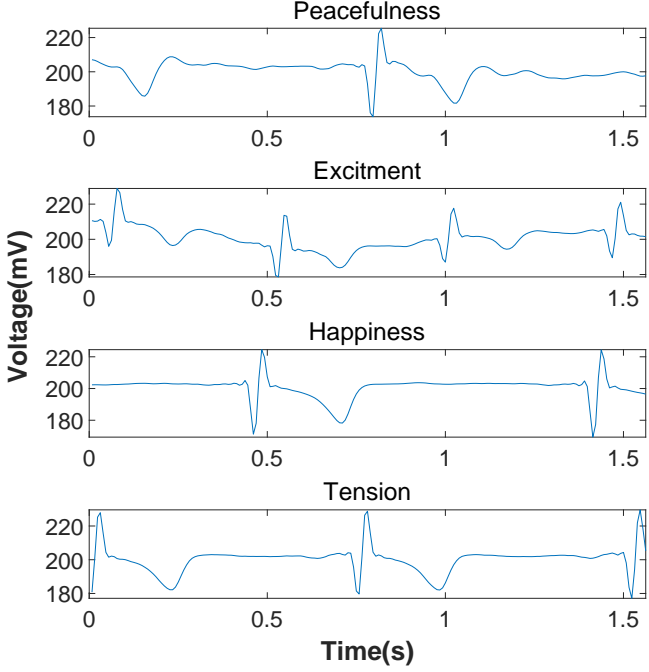}\caption{\textbf{The ECG segments with 200 points (1.56s) from four emotions}}
\label{f1}
\end{figure}

The DCT uses a sum of $N$ cosine functions at different frequencies to express finite data samples. It converts temporal signals into spectral signals. Eq.~(\ref{eq1}) defines the DCT formula for a data sequence $x(n)$, which is a Fourier transform without the conjugate portion.
\begin{eqnarray}
&y(k)=w(k)\sum_{n=1}^{N}x(n)\cos[\frac{\pi}{2N}(2n-1)(k-1)], \label{eq1}\\
& k=1,...,N, \nonumber   
\end{eqnarray}
where
\begin{eqnarray}
 w(k)=\protect\begin{cases}
\frac{1}{\sqrt{N}} & ,k=1\protect\\
\sqrt{\frac{2}{N}} & ,2\protect\leq k\protect\leq N
\protect\end{cases}
\end{eqnarray}
and $N$ is the length of the data sequence~\cite{dct_}.

\begin{figure}[htbp]
\centering
\includegraphics[width=\columnwidth,trim={2.7cm 6.4cm 4cm 6.5cm},clip]{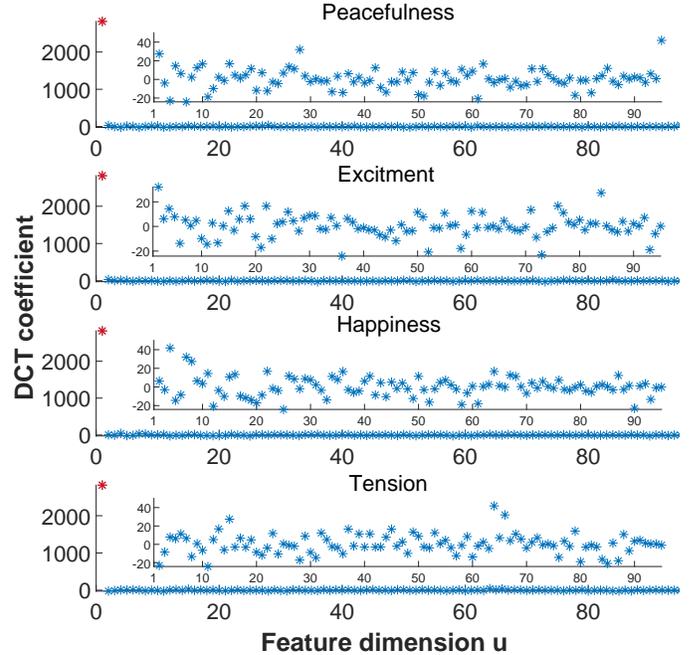}\caption{\textbf{The extracted features from ECG segments with dimension of 95}}
\label{f3}
\end{figure}

During DCT, data samples from each ECG segment are translated into the frequency domain, generating a series of DCT coefficients with length $N$. Then, the generated DCT coefficients are arranged in a decreasing order based on their absolute values. DCT coefficients with larger absolute values are treated as significant features which will be fed into the proposed X-GWO-SVM scheme. Descending DCT coefficients with dimension $u$ ($u \leq N$) can be selected as the extracted features.  
The determination of a proper dimension $u$ of extracted features will be discussed in Section~\ref{Classification results} by comparing classification performances at different values. Fig.~\ref{f3} shows corresponding extracted features with dimension $u=95$, i.e., coefficients with the largest 95 absolute values, from the aforementioned ECG segments (plotted in Fig.~\ref{f1}). It should be noticed that the first coefficient takes the highest energy (highlighted with red color), which stores the most significant features. To observe details on the rest coefficients, we zoom in the rest of coefficients (the $2_{nd}$ to $95_th$ coefficients) for each emotion. Compared to the original ECG signals, the four emotions are clearly differentiated between segments following feature extraction. 

\subsection{Exploitative and explorative grey wolf optimizer based support vector machine}
\label{X-GWO-SVM}
\begin{figure}[htbp]
\centering
\includegraphics[width=\columnwidth ,trim={6cm 3.9cm 6cm 0.5cm},clip]{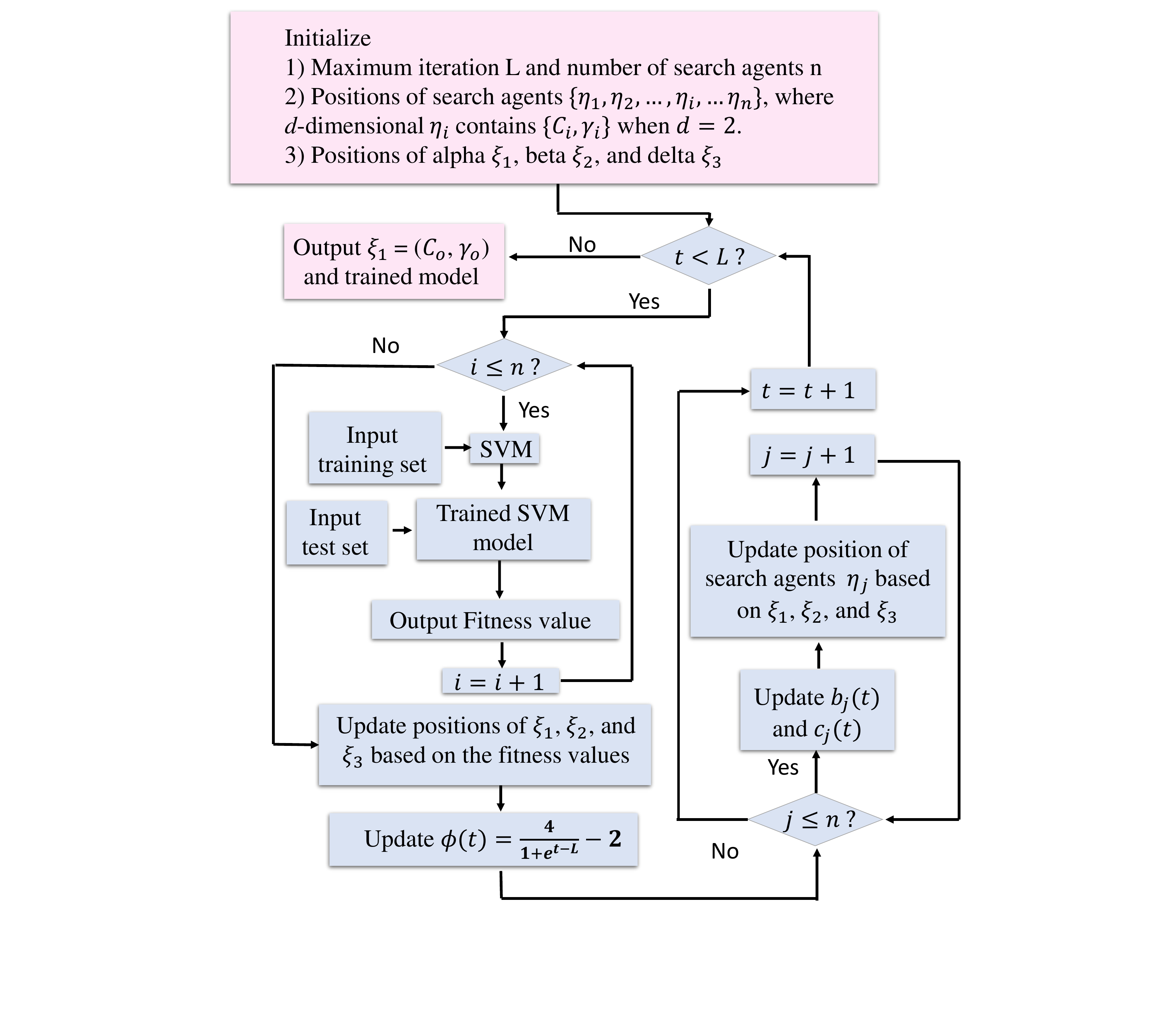}
\caption{\textbf{Flow chart of the X-GWO-SVM algorithm. Total number of search agents (wolves) is represented by $n$.} \textmd{$C_o$ and $\gamma_o$ are components stored in the final fittest solution $\mathbf{\xi}_{1}$.}}
\label{f6}
\end{figure}

For the first time, the X-GWO-SVM approach is proposed for ECG emotion identification in this work. The hyperparameter-free property of the proposed method provides a new way for radial basis function-based SVM (RBF-SVM) learning. 
In general, classifying the non-linearly separable data with RBF-SVM requires two hyperparameter which are considered---a penalty coefficient $C$ and a spacial parameter $\gamma$. The objective function of RBF-SVM with $C$ and $\gamma$ introduced is expressed in Eqs.~\eqref{eqC} and~\eqref{eqg}:
\begin{eqnarray}
    &\min\frac{1}{2}||\mathbf{w}||^2+ C \sum_{i=1}^P\epsilon_i, \label{eqC}\\ 
    &s.t. \quad  y_i(\mathbf{w}^T\Psi (\mathbf{x}_{i})+b)-1+\epsilon_i\geq 0 \nonumber  \\
    &\quad \forall i=1,2, ..., P,\nonumber \\
    & \mathbf{K}(\mathbf{x}_{i},\mathbf{x}_{j})=e^{-\gamma \parallel \mathbf{x}_{i}-\mathbf{x}_{j}\parallel^{2}} 
    = \Psi(\mathbf{x}_{i})^T\Psi(\mathbf{x}_{j}),\label{eqg}
\end{eqnarray}
where $P$ is the number of training samples; $\epsilon_i$ is a slack variable which is added to relax the constraints of linear SVM; $\mathbf{w}^T\Psi (\mathbf{x}_{i})+b$ is the decision function; $y_i$ is the class label; $\mathbf{x}_{i}$ is the sample; $C$ is the penalty parameter and it controls the trade-off between the size of the margin and the slack variable penalty; $\gamma$ is a spacial parameter which controls data distribution in a new feature space~\cite{pso-svm(2),tharwat2019parameter}. 
Obviously, hyperparameter ($C$ and $\gamma$) tuning for RBF-SVM is necessary but complex. Thus, the proposed method can internally learn hyperparameters by emphasizing the importance of the $\alpha$ wolf and non-linearly updating coefficient vectors used in GWO. Moreover, this method has higher recognition accuracy than the existing GWO-SVM and PSO-SVM techniques for ECG emotion recognition use, and it can effectively avoid the algorithm falling into a local solution by increasing the exploration ability and speed up the convergence ability by increasing the exploitation ability.

Fig.~\ref{f6} demonstrates our X-GWO-SVM method, which is inspired by the activity of grey wolves. There are 4 types of grey wolves, named alpha ($\alpha$), beta ($\beta$), delta ($\delta$), and omega ($\omega$), simulating the leadership hierarchy. These wolves continuously search for prey, the optimal solution in our case, and hunting (optimization) is guided by the fittest solution, second and third best solutions, $\alpha$, $\beta$ and $\delta$, respectively. The $\omega$ wolves follow these three wolves. A total number of search agents (wolves) is represented by $n$. 
$C_o$ and $\gamma_o$ are two elements of the searched optimal solution $\mathbf{\xi}_{1}$. The X-GWO-SVM method has 10 steps as described below.
\begin{enumerate}
    \item{\label{1}} The X-GWO-SVM related parameters are initialized, i.e., maximum iteration $L$; the number of search agents $n$; positions of $\alpha$ ($\mathbf{\xi}_1$), $\beta$ ($\mathbf{\xi}_2$) and $\delta$ ($\mathbf{\xi}_3$); positions of search agents (wolves) ${\mathbf{\eta}_1, \mathbf{\eta}_2, ..., \mathbf{\eta}_i, ..., \mathbf{\eta}_n}$. $\mathbf{\eta}_i\in \mathbb{R}^{d}$ and $\mathbf{\xi}_i\in \mathbb{R}^{d}$ are $d$-dimensional vectors. In this case, $d$ is equal to $2$, representing two optimal hyperparameters ($C$ and $\gamma$) required for search.  

    \item{\label{2}} If the current iteration time $t$ is less than the maximum iteration $L$, go to the subsequent steps; otherwise, proceed directly to step~\ref{17}).
     
    \item For each agent, train RBF-SVM with current position elements $\eta_i = (C_i, \gamma_i)$.
    \item Predict trained RBF-SVM with the test set for each agent and output its loss as a fitness value based on Eq.~\eqref{loss_eq}:
    \begin{equation}
    \label{loss_eq}
        loss(\eta_i) = \frac{1}{M}\sum_{i=1}^{M}(y_i-h_i)^2, 
    \end{equation} 
    where $M$ is the number of test samples and $h_i$ represents the predicted value for the $i_{th}$ test sample.
    \item Sort all fitness values in ascending order and assign positions which have the corresponding top three fitness values as $\mathbf{\xi}_{1}$, $\mathbf{\xi}_{2}$ and $\mathbf{\xi}_{3}$, respectively. The mathematical expressions are
    \begin{align}
    \label{eqn11}
     \mathbf{\xi}_{1}(t) &= \argmax_{\eta_i(t),i=1, ..., n}loss(\eta_i(t)),\\
    \label{eqn22}
     \mathbf{\xi}_{2}(t) &= \argmax_{\eta_i(t);\eta_i(t)\neq \mathbf{\xi}_{1}(t)}loss(\eta_i(t)),\\
    \label{eqn33}
     \mathbf{\xi}_{3}(t) &= \argmax_{\eta_i(t);\eta_i(t)\neq \mathbf{\xi}_{1}(t),\mathbf{\xi}_{2}(t)}loss(\eta_i(t)).
    \end{align}

    \item{\label{9}} Update exploration-exploitation regulation function $\phi(t)$ based on Eq.~\eqref{eqn5}:
    \begin{equation}
        \phi(t) = -2\frac{t-L+1}{-t+L}.
        \label{eqn5}
    \end{equation}

    \item For each search agent, update its position $\mathbf{\eta}_i$ based on following equations:
    
        \begin{align}
        \label{eqn3}
         \mathbf{\eta}_i(t+1) &= \frac{1}{4}\mathbf{\xi}_1(t)+\frac{1}{4}\sum_{i = 1}^3 [\mathbf{\xi}_i(t) \\
         &-b_i(t)\odot|c_i(t)\odot\mathbf{\xi}_i(t)-\mathbf{\eta}_i(t)|],\nonumber \\
        \label{eqn4}
        b_i(t) &= 2\phi(t)r_i(t)-\phi(t)\mathbf{1},
        \quad c_i(t) = 2s_i(t),
        \end{align}
    where $\odot$ and $|\cdot|$ represent Hadamard product operation and element wise absolute value operations, respectively; $t$ is the iteration number; $\mathbf{1}\in \mathbb{R}^{2}$ and its elements are all ones; $b_i\in \mathbb{R}^{2}$ and $c_i\in \mathbb{R}^{2}$ are coefficient vectors. The coefficients $r_i\in \mathbb{R}^{2}$ and $s_i\in \mathbb{R}^{2}$ are random vectors, where elements are in the range 0 to 1. 

    \item {\label{16}} Accumulate iterative time and go back to step~\ref{2}).
    \item{\label{17}} Output the optimal parameters $\xi_1=(C_o,\gamma_o)$ and the trained SVM model.
    \item Calculate the classification accuracy of the model based on the test set and end the X-GWO-SVM algorithm.    
\end{enumerate}

We demonstrate improvements of the proposed X-GWO-SVM algorithm with respect to its exploration and exploitation ability in following two subsections. 

\subsubsection{Exploration}

\begin{figure} [htbp]
    \centering
  \subfloat[\label{1a}]{%
       \includegraphics[width=\linewidth,trim={3cm 9cm 4cm 9.9cm},clip]{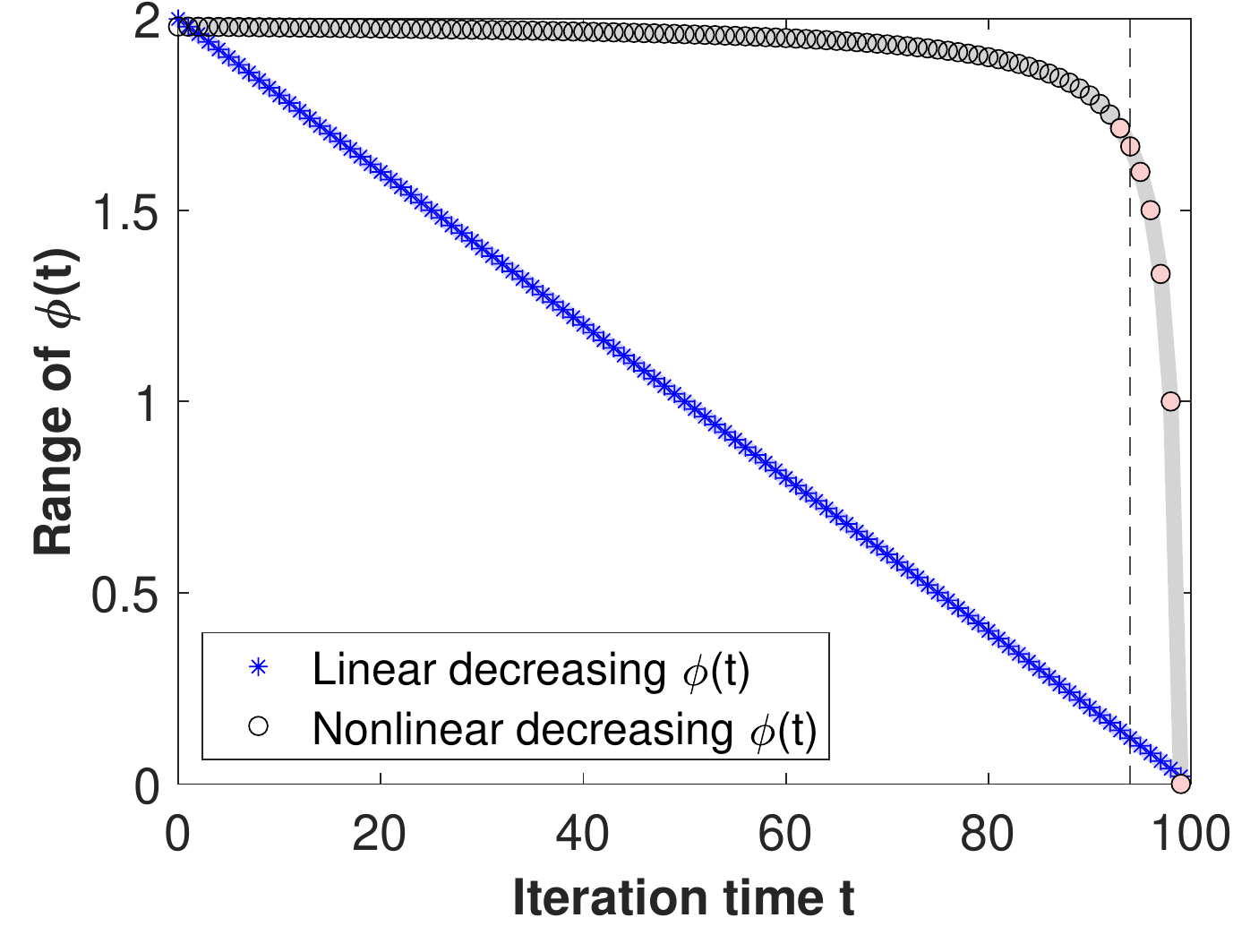}}
    \hfill
    \centering
  \subfloat[\label{1b}]{%
        \includegraphics[width=\linewidth,trim={3cm 9cm 4cm 10cm},clip]{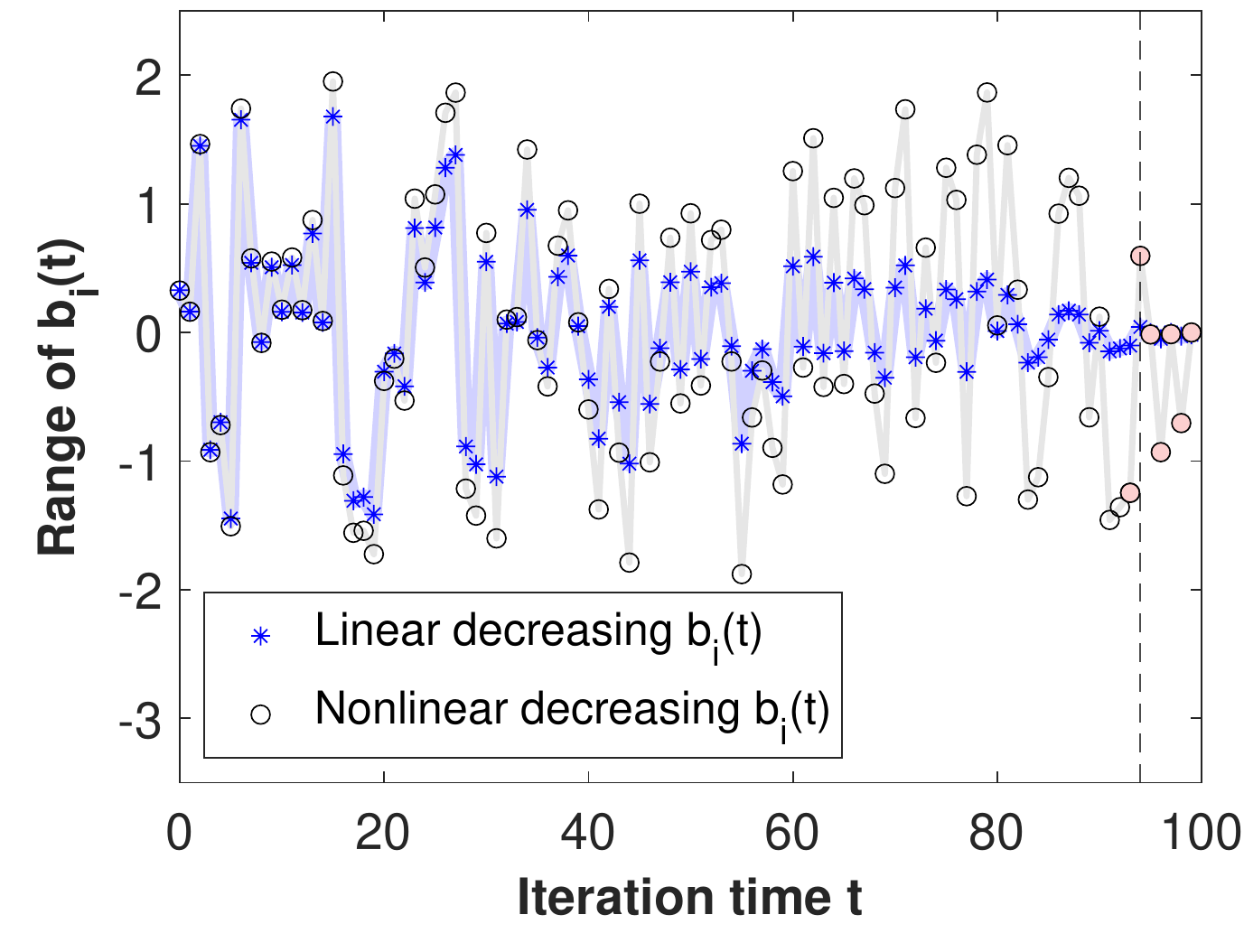}}
 
  \caption{\textbf{(a) Components of $\phi(t)$ linearly (blue circles) and non-linearly (black circles) decreased from 2 to 0 over the course of iterations. }
\textbf{(b) The corresponding variations for components of $b_i(t)$ based on the components of $\phi(t)$ linearly (blue circles) and non-linearly (black circles) decreased over the course of iterations.} 
\textmd{The maximum iteration time $L$ is set to 100 for both cases. 
The black dash line lies at the $94_{th}$ iteration. The red-filled circles aim to clearly indicate variations of $\phi(t)$ and $b_i(t)$ for the nonlinear case after the $94_{th}$ iteration.}
}
  \label{a_coeff} 
\end{figure}

Conventionally, components of $\phi(t)$ are linearly decreased from 2 to 0 over the course of iterations~\cite{mirjalili2014grey}, which models wolves approaching the prey. In our design, we set components of $\phi(t)$ non-linearly decrease from 2 to 0 with slower declining rate near 2 and faster declining rate near 0 (referring to Eq.~\eqref{eqn5}). Fig.~\ref{a_coeff} (a) demonstrates the components of $\phi(t)$ linearly (blue stars) and non-linearly (black circles) decreased from 2 to 0 over the course of iterations when the maximum iteration time $L$ is set to 100. Clearly, for the designed nonlinear decreasing method, we can observe that there is slow declining at the left side of the black dash line (iteration time = $94$), aiming to explore a larger range and increase exploration compared to the conventional linear way.

As discussed in~\cite{mirjalili2014grey}, $b_i(t)$
with random values greater than 1 or less than -1 is used to oblige the search agent to diverge from the prey, which emphasizes exploration and allows the X-GWO-SVM algorithm to search globally.
It should be noticed that the fluctuation range of $b_i(t)$ is also decreased under an effect of $\phi(t)$. Components of $b_i(t)$ are random values in the interval $[-\phi(t), \phi(t)]$, where components of $\phi(t)$ are non-linearly decreased from 2 to 0 over the course of iterations~\cite{mirjalili2014grey}. Fig.~\ref{a_coeff} (b) shows a variation of $b_i(t)$ when linear and nonlinear $\phi(t)$ are applied. The blue and grey shadows indicate variation trends for $b_i(t)$ when linear $\phi(t)$ and nonlinear $\phi(t)$ are applied, respectively.
Obviously, the value of $b_i(t)$ (black circles) for nonlinear $\phi(t)$ applied has a larger range compared with the value of $b_i(t)$ (blue stars) for linear $\phi(t)$ at the left side of the black dash line, i.e., $|b_i(t)|>\mathbf{1}$. In other words, the next search range for the fittest position in the nonlinear case smoothly attenuates before iteration reaches a threshold---94 in this figure, making sure a large exploration range.

\subsubsection{Exploitation (convergence)}

In~\cite{mirjalili2014grey}, when updating the positions, the weights for $\alpha$, $\beta$, and $\delta$ wolves are all the same.  While for our proposed approach, when updating the positions, we assign more weight to the $\alpha$ wolf (referring to Eq.~\eqref{eqn3}), which emphasizes the importance of the $\alpha$ wolf. Consequently, the fittest solution from the previous iteration can be retained and continually influences the subsequent updating step, ensuring a faster convergence.

We can observe from Fig.~\ref{a_coeff} (a) that, for the designed nonlinear decreasing method, there is a much faster decay at the right side of the black dash line. To clearly track the convergence of $\phi(t)$, red-filled circles are utilized for the nonlinear case after the $94_{th}$ iteration. The convergence tends to speed up as the iteration continuously increases, whereas, for the conventional linear decreasing method, the components of $\phi(t)$ just evenly decrease from 2 to 0.

As we discussed before, the value of $b_i(t)$ is influenced by the value of $\phi(t)$. Therefore, a similar phenomenon can be observed in Fig.~\ref{a_coeff} (b), where the value of $b_i(t)$ converges much faster than the linear case at the right side of the black dash line, i.e., $|b_i(t)|<\mathbf{1}$. 
In other words, the next search range for the fittest position in the nonlinear case dramatically decreases after iteration time reaches 94.

To sum up, the proposed X-GWO-SVM algorithm does not require hyperparameter tuning on SVM in order to get good accuracy. Additionally, it improves the way of updating position by involving the fittest position $\alpha$, which emphasizes the importance of the $\alpha$ wolf and keeps the effect of the fittest solution for the next iteration. We also enhance the ability of exploitation by nonlinearly decreasing the value of $\phi(t)$. The algorithm improves its global search ability by increasing the exploration ability and speeds up the convergence ability by increasing the exploitation ability.

\subsection{Measurements}

The classification performance of various methods can be evaluated by standard statistical measurements: accuracy (ACC) and F1-score (F1), defined as
\begin{align}
\text{ACC} &= \frac{\text{TP}+\text{TN}}{\text{TP}+\text{FP}+\text{TN}+\text{FN}}, \\
\text{F1} &= 2\frac{\text{PRE}\times\text{REC}}{\text{PRE}+\text{REC}},\\
\text{REC} &= \frac{\text{TP}}{\text{TP}+\text{FN}},
\quad \text{PRE} = \frac{\text{TP}}{\text{TP}+\text{FP}},
\label{eq:8}
\end{align}
where TP (true positive) is the number of samples correctly predicted as the current class; TN (true negative) means the number of correctly predicted as other classes; FP (false positive) indicates the number of samples incorrectly detected as the current class; FN (false negative) denotes the number of samples incorrectly detected as other classes.
Accuracy is the general measurement of the correctly predicted ratio of the total testing samples for each dataset, indicating the method’s capability to classify emotions correctly. The F1-score, on the other hand, more accurately captures the ideal model for the unbalanced class distribution. The goal is to maximize these two measures as representations of effective models.

\section{Results}
\label{Classification results}

\subsection{Feature dimension selection}
As aforementioned in Section~\ref{Discrete cosine transform}, determination of a proper number of extracted features is necessary. Therefore, we perform feature importance selection in the range of 20 to 135 with step size 5 under the X-GWO-SVM method for iRealcare dataset. Each simulation result is repeated 10 times for random selection of training and test samples. 

Fig.~\ref{f7} illustrates the accuracy versus the dimension of the feature under the X-GWO-SVM algorithm with 10-fold cross-validation for the iRealcare dataset. Clearly, the recognition accuracy displays the tendency to rise up at the beginning and decline in late. The highest mean accuracy is 93.37\% located at the feature dimension equal to 95. Moreover, its corresponding box plot (filled with orange color) has relatively low variance, indicating the stability of this feature dimension. After getting the most discriminative result with feature dimension 95, we apply it to other comparison methods.

\begin{figure}[htbp]
\centering
\includegraphics[width=\columnwidth ,trim={2.5cm 8.5cm 3.5cm 8.5cm},clip]{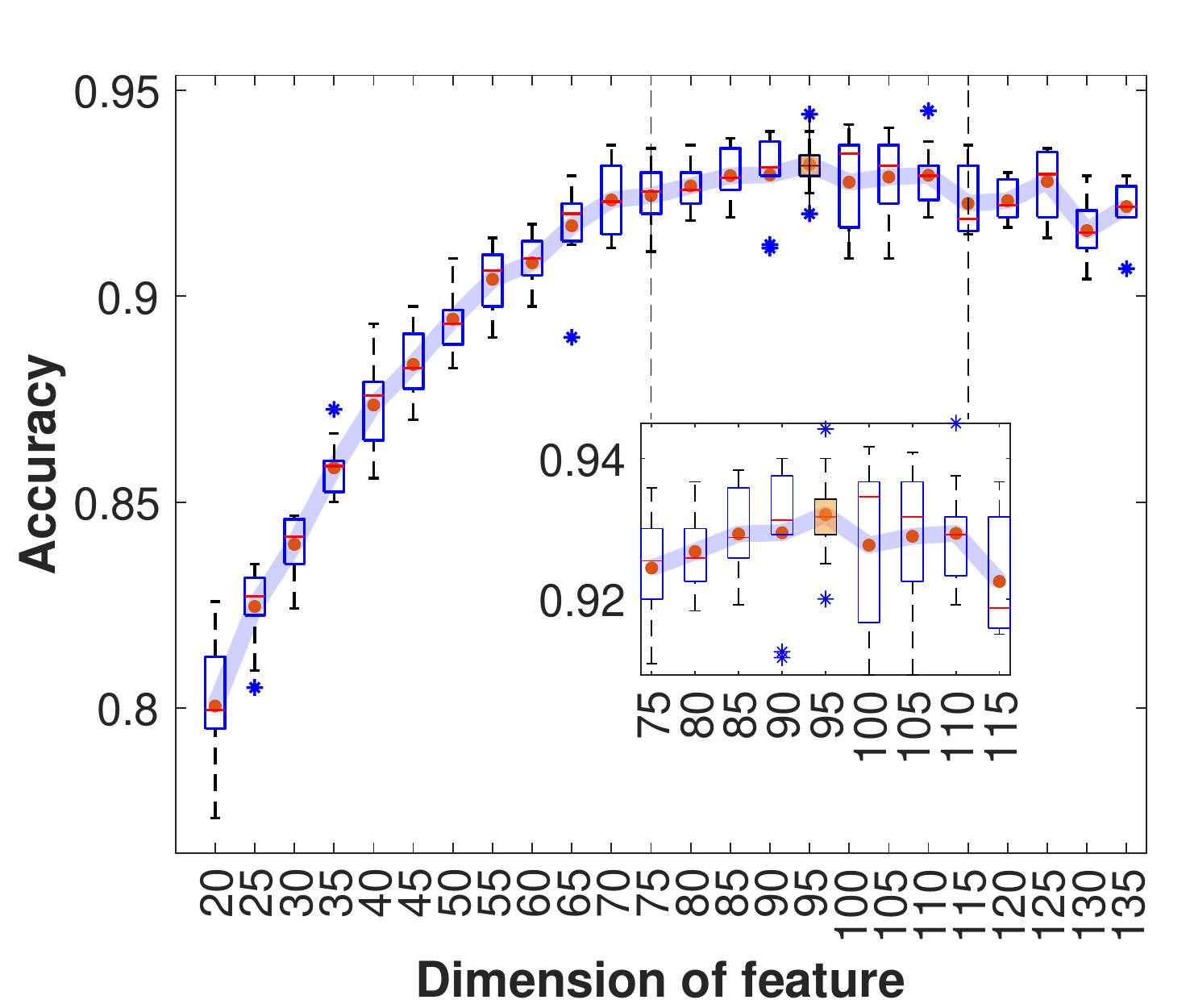}
\caption{\textbf{The accuracy versus the dimension of the feature under the X-GWO-SVM algorithm with 10-cross validation for the iRealcare dataset.}
\textmd{The blue shadow indicates a trend for mean accuracy values of different dimensions of features under the X-GWO-SVM algorithm with 10-cross validation.
Box plot is employed with the box top and bottom denoting the 75th and 25th percentiles respectively for the results of 10-cross validation; The red straight line inside the box denotes the median value, while the red dot denotes the mean value; The blue star denotes the outlier value. The most discriminative result is filled with orange color.}}
\label{f7}
\end{figure}

\subsection{Exploration-exploitation regulation function selection}
\label{feature_selection}

\begin{figure}[htbp]
\centering
\includegraphics[width=\columnwidth ,trim={3.2cm 9cm 4cm 9.5cm},clip]{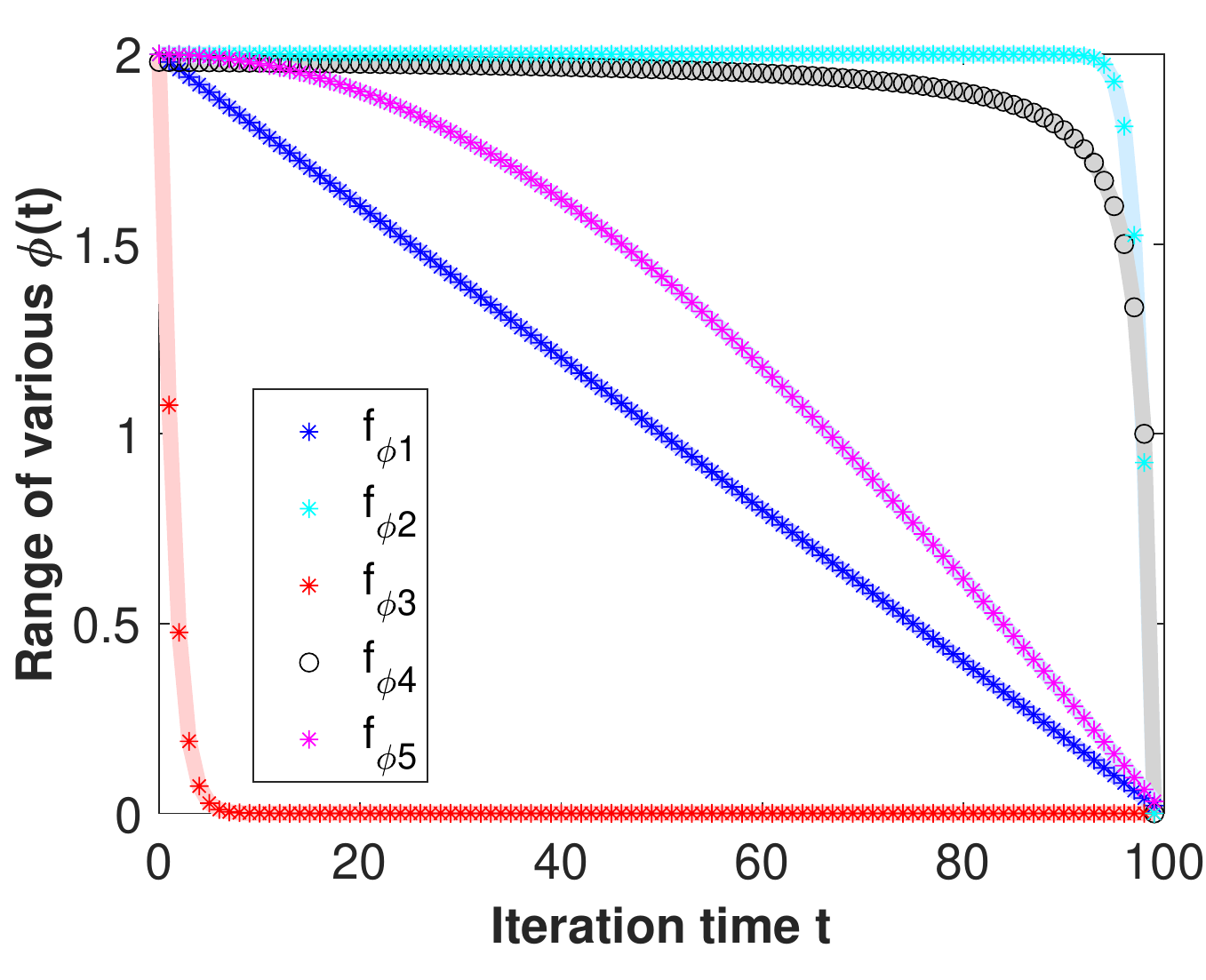}\caption{\textbf{Variations of components of $\phi(t)$ for different exploration-exploitation regulation functions.} }
\label{f5}
\end{figure}

As we demonstrated the significance of exploration-exploitation regulation function $\phi(t)$ in Section~\ref{X-GWO-SVM}, various exploration-exploitation regulation functions are used in our experiments here to demonstrate that our choice of $\phi(t)$ used in the X-GWO-SVM algorithm is the best. Expressions on them are shown in Eqs.~\eqref{eqn6} to~\eqref{eqn10} and these exploration-exploitation regulation functions are plotted in Fig~\ref{f5}. It should be noticed that the conventional linear exploration-exploitation regulation function, a benchmark, is expressed in Eq.~\eqref{eqn6}. Additionally, the one we proposed in the X-GWO-SVM in Eq.~\eqref{eqn5} is rewritten as $f_{\phi 4}(t)$ in Eq.~\eqref{eqn9}.
\begin{align}
\label{eqn6}
f_{\phi 1}(t)&= 2-\frac{2t}{L},\\
\label{eqn7}
f_{\phi 2}(t)&= \frac{4}{1+e^{ t-L}}-2,\\
\label{eqn8}
f_{\phi 3}(t)&= \frac{-4}{1+e^{-t}}+4,\\
\label{eqn9}
f_{\phi 4}(t)&= -2\frac{t-L+1}{-t+L} = \phi(t),\\
\label{eqn10}
f_{\phi 5}(t)&= 2\cos(\frac{\pi}{2tL}).
\end{align}
Based on Fig.~\ref{f5}, we can observe that both $f_{\phi 2}$ and $f_{\phi 3}$ are deformed from the Sigmoid function, which dramatically decrease from 2 to 0 at the beginning and the end of the iteration, respectively.  
The function $f_{\phi 4}$ and function $f_{\phi 5}$ successively alleviate this decreasing trend on a basis of function $\frac{1}{x}$ and $\cos$, respectively.

To evaluate the effects of exploration-exploitation regulation function, we apply 10-fold cross-validation to the proposed X-GWO-SVM, varying exploration-exploitation regulation functions based on the aforementioned five functions. The evaluated results on exploration-exploitation regulation functions are shown in Table~\ref{tab2} for iRealcare dataset and Table~\ref{tabw} for WESAD dataset. Clearly, for iRealcare dataset, X-GWO-SVM combined with $f_{\phi 4}$ has the highest accuracy (93.37\%) and F1-score (93.38\%) among others. Moreover, the lowest variance and pretty low training time indicate its stability with low computation time. A similar conclusion can be derived for results on the WESAD dataset, where the highest accuracy (95.93\%) and F1-score (95.56\%) are from the combination of X-GWO-SVM with $f_{\phi 4}$. To this end, we have determined the optimal feature dimension---95, and exploration-exploitation regulation function---$f_{\phi 4}$. Therefore, later evaluations of the proposed X-GWO-SVM are based on these two settings.

\begin{table}[h]
\caption{Results of X-GWO-SVM involved with five exploration-exploitation regulation functions on iRealcare dataset.}
\begin{adjustbox}{width=3.6 in,center}
\centering
\begin{tabular}{ccccc}
\hline
\hline	

Exploration-exploitation regulation function	&	Mean(ACC)	&	Var(ACC)	&	Mean(F1)/\%		&	Training time/s	\\
\hline
$f_{\phi 1}$	&	92.90	&	9.66E-05	&	92.91	&	429.98	\\
$f_{\phi 2}$	&	93.05	&	6.10E-05	&	93.08	&	402.80	\\
$f_{\phi 3}$	&	83.26	&	1.40E-02	&	83.50	&	509.66	\\
$\mathbf{f_{\phi 4}}$	&	\textbf{93.37}	&	\textbf{2.84E-05}	&	\textbf{93.38}	&	\textbf{380.16}	\\
$f_{\phi 5}$	&	92.93	&	3.31E-05	&	92.94	&	368.14	\\

\hline	
\hline
\end{tabular}
\label{tab2}
\end{adjustbox}
\end{table}

\begin{table}[h]
\caption{Results of X-GWO-SVM involved with five exploration-exploitation regulation functions on WESAD dataset.}
\begin{adjustbox}{width=3.6 in,center}
\centering
\begin{tabular}{ccccc}
\hline
\hline	

Exploration-exploitation regulation function	&	Mean(ACC)	&	Var(ACC)	&	Mean(F1)/\%		&	Training time/s	\\
$f_{\phi 1}$	&	95.29	&	1.62E-04	&	95.07& 856.31	\\
$f_{\phi 2}$	&	95.79	&	7.48E-05	&	95.44&828.31	\\
$f_{\phi 3}$	&	95.29	&	1.62E-04	&	95.07&825.14	\\
$\mathbf{f_{\phi 4}}$	&	\textbf{95.93}	&	\textbf{5.61E-05}	& \textbf{95.56} &	\textbf{813.58}	\\
$f_{\phi 5}$	&	95.79	&	7.48E-05	&	95.44&835.40	\\

\hline	
\hline
\end{tabular}
\label{tabw}
\end{adjustbox}
\end{table}

\subsection{Classification Performance of Proposed Model}

\subsubsection{Classification Performance for of iRealcare dataset}

One may suspect that only one of the improvements on X-GWO-SVM can achieve a considerable performance. Thus, we investigate the other three methods: 1) using the GWO-SVM method, where none of the improvement on GWO is applied; 2) using the nonlinear $\phi(t)$ based grey wolf optimizer (N-GWO-SVM) method, where only the nonlinearly decreasing value of $\phi(t)$ is used; 3) using PSO-SVM method, where the conventional PSO algorithm is used for searching optimal hyperparameters.

Tables~\ref{tab3} to~\ref{tab5} show the classification performance for the hyperparameter optimizer-based techniques stated above.
The following metrics are reported: accuracy, F1-score, variation of accuracy, and training duration of the schemes. All of them are calculated from 10 repeated classification trials for each scheme (rows in Tabs.~\ref{tab3} to ~\ref{tab5}). 

\begin{table}[h]
\caption{The mean ACC of four emotions with hyperparameter optimizer based schemes evaluated on iRealcare dataset.}
\begin{adjustbox}{width=3.2 in,center}
\centering
\begin{tabular}{ccccc}
\hline
\hline	
Scheme	&	Peacefulness/\%	&	Excitement/\%	&	Happiness/\%	&	Tension/\%	\\
\hline
PSO-SVM	&	83.20	&	84.87	&	\textbf{95.40}	&	91.13\\
GWO-SVM	&	84.07	&	97.67	&	94.73	&	91.40\\
N-GWO-SVM	&	84.60	&	97.93	&	94.33	&	91.27\\
\textbf{X-GWO-SVM}	&	\textbf{86.03}	&	\textbf{98.03}	&	95.07	&	\textbf{94.33}\\

\hline	
\hline
\end{tabular}
\label{tab3}
\end{adjustbox}
\end{table}

Table~\ref{tab3} shows that GWO-SVM performs significantly better than PSO-SVM for peacefulness (84.07\% vs. 83.20\%), excitement (97.67\% vs. 84.87\%), and tension (94.73\% vs. 91.13\%), but N-GWO-SVM only slightly improved performance on peacefulness (84.60\%) and excitement (97.93\%).
Except for a slightly lower performance on happiness compared to the PSO-SVM scheme (95.07\% vs 95.40\%), the proposed X-GWO-SVM scheme provides a significant performance boost over others. It has the highest accuracy for peacefulness, excitement, and tension of 86.03\%, 98.03\%, and 94.33\%, respectively. 
Table~\ref{tab_f1} presents similar results for the mean F1 score. The proposed X-GWO-SVM scheme provides a significant performance boost over others. 

\begin{table}[h]
\caption{The mean F1 of four emotions with hyperparameter optimizer based schemes evaluated on iRealcare dataset.}
\begin{adjustbox}{width=3.6 in,center}
\centering
\begin{tabular}{ccccc}
\hline
\hline	
Scheme	&	Peacefulness/\%	&	Excitement/\%	&	Happiness/\%	&	Tension/\%	\\
\hline
PSO-SVM	&	85.29	&	82.46	&	91.07	&	94.45\\
GWO-SVM	&	84.19	&	97.87	&	90.77	&	93.57\\
N-GWO-SVM	&	84.59	&	97.99	&	91.03&	93.36\\
\textbf{X-GWO-SVM}	&	\textbf{87.73}	&	\textbf{98.31}	&	\textbf{91.23}	&	\textbf{96.24}\\

\hline	
\hline
\end{tabular}
\label{tab_f1}
\end{adjustbox}
\end{table}

\begin{table}[h]
\caption{The mean variance of ACC for four emotions with  hyperparameter optimizer based schemes evaluated on iRealcare dataset.}
\begin{adjustbox}{width=3.in,center}
\centering
\begin{tabular}{ccccc}
\hline
\hline	
Scheme	&	Peacefulness	&	Excitement	&	Happiness	&	Tension	\\
\hline
PSO-SVM	&	\textbf{6.00E-04}	&	7.17E-02	&	3.00E-04	&	3.00E-04	\\
GWO-SVM	&	5.00E-04	&	2.00E-04	&	4.00E-04	&	1.10E-03		\\
N-GWO-SVM	&	1.60E-03	&	2.00E-04	&	1.4E-04	&	3.00E-04		\\
\textbf{X-GWO-SVM}	&	6.01E-04	&	\textbf{6.04E-05}	&	\textbf{1.38E-04}	&	\textbf{7.65E-05}\\
\hline	
\hline
\end{tabular}
\label{tab4}
\end{adjustbox}
\end{table}

The variance result in Table~\ref{tab4} suggests a similar conclusion.
Compared to the conventional PSO-SVM scheme or GWO-SVM scheme, except for the variance on peacefulness (6.01E-04 vs 6.00E-04), our proposed method also shows the lowest variance of accuracy on excitement (6.04E-05), happiness (1.38E-04) and tension (7.65E-05), indicating its stability.

\begin{table}[h]
\caption{The mean values of ACC, F1, variance and training time for four emotions with hyperparameter optimizer based schemes evaluated on iRealcare dataset.}
\begin{adjustbox}{width=3.2 in,center}
\centering
\begin{tabular}{ccccc}
\hline
\hline	
Scheme	&	Mean(ACC)/\%	&Mean(F1)/\%&	Mean(Var) &	Training time/s	\\
\hline
PSO-SVM	&	88.65	& 88.32&	1.82E-02	&	8824.80	\\
GWO-SVM	&	91.97	&	91.60&5.50E-04	&	446.24	\\
N-GWO-SVM	&	92.03	&91.74&	5.25E-04	&	446.42	\\
\textbf{X-GWO-SVM}	&	\textbf{93.37}	&\textbf{93.38}&	\textbf{2.19E-04}	&	\textbf{380.16}	\\

\hline	
\hline
\end{tabular}
\label{tab5}
\end{adjustbox}
\end{table}

Table ~\ref{tab5} shows the mean accuracy, the mean variance and the mean training time of 10-fold cross-validation results. Note that, the proposed X-GWO-SVM scheme has the highest mean of class accuracy (93.37\%), the highest mean of class f1-score (93.38\%), the lowest mean of class variance (2.19E-04), and the shortest hyperparameter training time (380.16s). The results demonstrated its high reliability, stability, and efficiency.

\begin{table}[h]
\caption{The mean ACC, mean F1 and variance of nonparametric classification methods and the proposed method evaluated on iRealcare dataset.}
\begin{adjustbox}{width=2.5 in,center}
\centering
\begin{tabular}{cccc}
\hline
\hline	
Algorithm	&	Mean(ACC)/\% &Mean(F1)/\%	&	Mean(Var)	\\
\hline
RF	&	81.71	&82.35	&\textbf{1.56E-04}	\\
K-NN	&	82.48&81.22	&	7.35E-04	\\
\textbf{X-GWO-SVM}	&	\textbf{93.37}&\textbf{93.38}	&	2.19E-04\\

\hline	
\hline
\end{tabular}
\label{tab6}
\end{adjustbox}
\end{table}

To demonstrate the the significance of our X-GWO-SVM scheme, we also examine other nonparametric classification methods using similar features on the iRealcare dataset, such as RF and K-NN. The results are shown in Table~\ref{tab6}. All of them are calculated from 10 repeated classification trials for each scheme. It is apparent from the results that the X-GWO-SVM scheme has the highest accuracy performance. 
Actually, RF is more stable than the X-GWO-SVM, while its accuracy and F1-score are much lower than the proposed scheme (81.71\% vs 93.37\%, 82.35\% vs 93.38\%, respectively).
In addition, the X-GWO-SVM completely outperforms K-NN in terms of reliability and stability.
Overall, the proposed X-GWO-SVM strategy is more stable and more effective at achieving high mean accuracy on the iRealcare dataset than the existing methods.

\subsubsection{Classification Performance for WESAD dataset}
To demonstrate the reliability and stability of the proposed X-GWO-SVM method, we further examine it on the WESAD dataset in terms of accuracy and  F1-score, compared with other existing methods. 
By applying the feature dimension selection in the range of 4000 to 10000 with step size 500 under the X-GWO-SVM approach as we described in Section~\ref{feature_selection}, the best feature dimension is found to be 5000.

Fig.~\ref{f7} illustrates the accuracy versus the dimension of the feature
under the X-GWO-SVM algorithm with 10-cross validation for
the WESAD dataset. Clearly, the recognition accuracy has a similar pattern to Fig.~\ref{f5}, in which it increases initially and decreases afterwards. The highest
mean accuracy is 95.93\% located at the feature dimension equal to 5000. The corresponding box plot has the highest accuracy---96.30\%, the lowest accuracy---94.44\%, and the mean F1-score---95.56\%. 
Moreover, its corresponding box plot also has a relatively low mean variance (5.49E-05), indicating the stability of this feature dimension. The mean training time is 113.07s, in this case.

Table~\ref{tab8} presents a comparison between the proposed classification scheme and the state-of-the-art methods published for single channel ECG-based emotion recognition methods on the WESAD dataset.
The same testing dataset ensures that the comparison is persuasive and feasible. 
It can be observed from Table~\ref{tab8} that the proposed classification approach outperforms all simple machine learning methods, such as RF, K-NN, linear discriminant analysis, and decision tree. 
Though slightly inferior to that of self-supervised CNN, the proposed X-GWO-SVM technique exhibits comparable classification performance among deep neural networks.
However, considering the computation complexity, the proposed method is much simpler and more efficient than the self-supervised CNN. Our algorithm has successfully been loaded into a lightweight embedded system with a prediction time of 2.659ms per 200-points iRealcare sample and 4.648ms per 14000-points WESAD sample. Details on these results will be discussed in Section~\ref{efficiency}.
Overall, the proposed X-GWO-SVM method achieves comparable accuracy and F1-score (Fig.~\ref{f7} and Table~\ref{tab8} ) among neural network-based deep learning classifiers on the WESAD dataset and has an overwhelming performance on other existing techniques.

\begin{figure}[htbp]
\centering
\includegraphics[width=\columnwidth ,trim={3cm 10.3cm 4.3cm 10.5cm},clip]{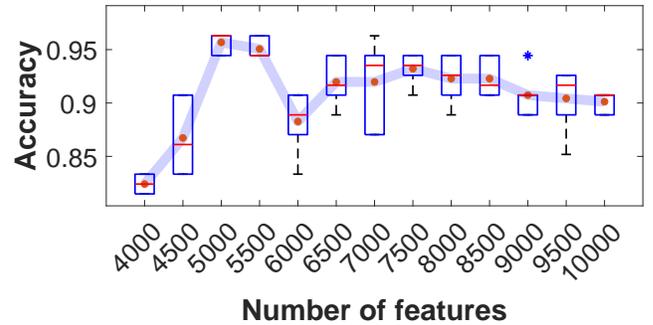}\caption{\textbf{The accuracy versus the dimension of the feature under the X-GWO-SVM-SVM algorithm with 10-cross validation on WESAD
dataset.} }
\label{f7}
\end{figure}

\begin{table}[h]
\caption{Comparison of various single channel ECG-based emotion recognition methods on WESAD dataset}
\begin{adjustbox}{width=\columnwidth,center}

\centering
\begin{tabular}{ccccc}
\hline
\hline

Reference	&	Year	&	Method	&	ACC/\%	&	F1/\%	\\
\hline
	&		&	RF	&	82.78	&	79.64	\\
	&		&	K-NN	&	79.19	&	75.39	\\
Schmidt et al. ~\cite{schmidt2018introducing}	&	2018	&	Linear discriminant analysis	&	85.44	&	81.31	\\
	&		&	AdaBoost Decision Tree	&	83.37	&	80.20	\\
	&		&	Decision Tree	&	80.17	&	77.01	\\
	\hline

Lin et al.~\cite{lin2019explainable}	&	2019	&	CNN	&	83.00	&	81.00	\\
\hline
Sarkar et al.~\cite{sarkar2020self}	&	2020	&	Self-supervised CNN	&	96.90	&	96.30	\\
	&		&	Fully-supervised CNN	&	93.20	&	91.20	\\
	\hline
\textbf{Proposed work}	&	——	&	\textbf{X-GWO-SVM}	&	\textbf{95.93}	&	\textbf{95.56}	\\

\hline
\hline
\end{tabular}
\label{tab8}
\end{adjustbox}
\end{table}

\section{Discussion}
\label{discussion}
The X-GWO-SVM algorithm, for the first time, is proposed and also the first time used in single channel ECG-based emotion recognition. By designing a suitable exploration-exploitation regulation function and updating technique, we are able to increase the exploration ability and exploitation ability with the proposed approach.
Two ECG datasets are used: one raw self-collected iRealcare dataset and one credible WESAD dataset. 
The X-GWO-SVM technique effectively avoids the algorithm from falling into a local solution; hence, it has a greater recognition accuracy than the existing GWO-SVM and PSO-SVM techniques for ECG emotion recognition.

The algorithm enables accurate, stable, and efficient emotion recognition based on single-channel ECG-based signals, which fills a gap for GWO-SVM research on ECG-based emotion recognition and also has the potential
for clinical use.

\subsection{Evaluation of datasets}
Despite the restricted number of subjects in the iRealcare dataset, the number of samples for each subject is sufficient since the time of data collection for each emotion is sufficient. 
It is true that the WESAD dataset contains a larger number of subjects; however, the sample length required for this dataset to achieve high accuracy, which is 14000, drastically reduces the actual number of samples, for example, 9 samples for each subject on amusement, 15-18 samples for each subject on stress, and 28-29 samples for each subject on the baseline.
On the contrary, the sample length required for the iRealcare dataset to achieve high accuracy, which is only 200. Therefore, the number of samples for the iRealcare dataset is much larger than the one in the WESAD dataset. 

The reason for the caused aforementioned situation might come from the way of giving external stimulus and recording data. For the iRealcare dataset, ECG signals for happiness, tension, and excitement are recorded when subjects watch comedies, watch thriller movies and do exercises, respectively. It should be noticed that emotions normally instantaneously occur and hold for a short period. Therefore, we only record the period that subjects are actually in that emotion condition and ignore the transition period. Clearly, the definition of different emotions under this external stimulus setting is clear and subjects are easy to get into a specific emotion. 
However, for the WESAD dataset, amusement condition signals are collected when subjects watch funny video clips; stress condition signals are collected when subjects are asked to provide public speaking and mental arithmetic tasks; baseline condition signals are collected when subjects sit/stand at a table and read magazines. In fact, subjects tend to take some time to transfer from one emotion condition to the other. However, such a transition period is also recorded in the WESAD dataset. Thus, the sample length need to be long enough to make sure not just the transition period is included. The shorter the time, the more probable it is that only transitional periods will be included in the sample.

To sum up, in spite of the fact that the iRealcare dataset has limited subjects, the actual number of samples is much larger than the one in the WESAD dataset. Moreover, due to the exclusive emotion transition period for the WESAD dataset, the selection of sample length for the iRealcare dataset is more flexible than the WESAD dataset. We use the widely-used WESAD dataset as a benchmark for further comparison to validate our proposed X-GWO-SVM algorithm.

\subsection{Evaluation of exploration-exploitation regulation function selection}
We study the impact of exploration-exploitation regulation functions on emotion recognition performance. For this study, we select possible base functions that control the significance of each exploration-exploitation regulation function as listed in Eqs.~\eqref{eqn6} to~\eqref{eqn10}. Table~\ref{tab2} and Table~\ref{tabw} show the emotion recognition performance for five exploration-exploitation regulation functions on the iRealcare dataset and WESAD dataset, respectively. This analysis provides in-depth insight into the effect of the exploration-exploitation regulation functions associated with the emotion recognition outcome. Furthermore, this analysis helps us narrow down the most suitable exploration-exploitation regulation function in order to achieve the best performance.

As we mentioned in Section~\ref{X-GWO-SVM}, the declining rate of the exploration-exploitation regulation function at the beginning and end with respect to the iteration time represents the exploration and exploitation ability of the proposed X-GWO-SVM.
From Table~\ref{tab2} and Table~\ref{tabw}, we notice that for the exploration-exploitation regulation function $f_{\phi 3}$, where the function is formed on a basis of the sigmoid function, the model performance on emotion recognition is poor since it is under-explored and under-exploited. However, for those exploration-exploitation regulation functions lying above the benchmark function $f_{\phi 1}$, the model shows significantly better performance.
Interestingly, the performance drops when exploration-exploitation regulation functions decline too fast ($f_{\phi 2}$) or too slow ($f_{\phi 5}$).
The function $f_{\phi 4}$ gives the highest performance for emotion recognition compared to others since it has the most suitable diverging and converging performance to the X-GWO-SVM algorithm. Moreover, the fact that $f_{\phi 4}$ outperformed other functions for both datasets is also indicative of its stability.

In summary, the analysis above shows that for all the exploration-exploitation regulation functions, when the declining rate of the function at the beginning is too large or too small, for example, $f_{\phi 3}$ or $f_{\phi 2}$, emotion recognition accuracy drops due to the under-exploration or over-exploration. 
This results in the X-GWO-SVM more easily falling into local solutions.
Similarly, when the declining rate of the function at the end is too large or too small, $f_{\phi 2}$ or $f_{\phi 3}$, the performance also drops due to the under-exploitation or over-exploitation in such cases becomes too difficult for the algorithm to properly find the global solution. Hence, we conclude that there is a trade-off between exploration and exploitation for the exploration-exploitation regulation functions associated with the proposed X-GWO-SVM algorithm, for which the proper exploration-exploitation regulation function $f_{\phi 4}$ is applied resulting in avoiding falling into local solutions.

\subsection{Evaluation of reliability, stability and efficiency of X-GWO-SVM}

This section discusses the performance of X-GWO-SVM for emotion recognition in terms of reliability, stability, and efficiency.
\subsubsection{Reliability}
\label{Reliability}
In our work, when only fusing exploration-exploitation regulation function $f_{\phi 4}$ with GWO-SVM, i.e., N-GWO-SVM, the classification accuracy and F1-score get improved (referring to Table~\ref{tab3}, Table~\ref{tab_f1} and Table~\ref{tab5}). This improvement indicates that involving a nonlinear exploration-exploitation regulation function can improve the recognition performance. Similarly, the classification accuracy and F1-score get further enhanced when the importance of the $\alpha$ wolf is emphasized, i.e., X-GWO-SVM, which shows the equivalent importance of the improvement on the X-GWO-SVM. 

The classification performance of X-GWO-SVM is superior to the classification performance of other hyperparameter optimizer-based systems, such as PSO-SVM and GWO-SVM. This indicates the high reliability of the proposed algorithm over existing common hyperparameter optimizer-based schemes. Furthermore, the X-GWO-SVM has the highest accuracy and F1-score among simple machine learning methods, such as RF, K-NN, decision tree, and linear discriminant analysis (Table~\ref{tab6} and Table~\ref{tab8}). Our analysis indicates that the X-GWO-SVM is more effective than simple machine learning approaches at avoiding local solutions. For the deep learning neural networks, such as CNN, the X-GWO-SVM can still outperform them except for a more complex single---self-supervised CNN~\cite{sarkar2020self}. Though the accuracy and F1-score of the X-GWO-SVM are slightly lower than the one from the self-supervised CNN, considering the efficiency, which will be discussed in Section~\ref{efficiency}, our algorithm is still competitive. 

\subsubsection{Stability}
The variance of the proposed method and existing works is computed to evaluate the stability of the methods. All simulation results are applied with 10-fold cross-validation. Similar to the discussion in Section~\ref{Reliability}, the X-GWO-SVM is the most stable algorithm among existing common hyperparameter optimizer-based schemes. Besides, similar results on both the iRealcare dataset and the WESAD dataset also indicate the stability of the proposed method. 

\subsubsection{Efficiency}
\label{efficiency}

Our works are implemented through both MATLAB version R2021b and Python 3.7 for feature extraction, model training, and prediction. For MATLAB, the computation is performed on a laptop with 11th Gen Intel(R) Core(TM) i7-11800H (2.2GHz and 32GB of RAM). The computation time for classifying a 200-points (1.56s) iRealcare sample and a 14000-points (20s) WESAD sample roughly spends 0.355ms and 0.778ms, respectively, using our proposed method. For Python, the computation is performed in JETSON NANO with Quad-core ARM Cortex-A57 MPCore Processor (1.43GHz and 4GB of RAM). The computation time for classifying a 200-points (1.56s) iRealcare sample and a 14000-points (20s) WESAD sample roughly spends 2.659ms and 4.648ms, respectively, using our proposed method.

Compared with the self-supervised CNN, a deep neural network, the proposed X-GWO-SVM is much simpler. The two-step self-supervised architecture involves deep convolutional blocks and several fully connected layers in~\cite{sarkar2020self}, which may not be realized in lightweight embedded systems. Whereas, our algorithm has successfully been loaded into JETSON NANO, an embedded system-on-module and developer kit with a prediction time of 2.659ms per 200-points iRealcare sample and 4.648ms per 14000-points WESAD sample. This provides a way to embed an ECG patch with the proposed algorithm, achieving edge computing for emotion recognition on ECG signals.

Moreover, the X-GWO-SVM is the most efficient algorithm among existing common hyperparameter optimizer-based schemes, which is evaluated by the training time. The other interesting point that can be found in Table~\ref{tab5} is that all GWO-SVM-based techniques take shorter training time than the PSO-SVM work, which is compatible with \cite{Emary2015}'s conclusion.

\subsection{Limitations and future directions}
The possible limitation of the current study would be that we only
investigate four exploration-exploitation regulation functions, i.e., $f_{\phi 2}$, $f_{\phi 3}$, $f_{\phi 4}$, and $f_{\phi 5}$. Moreover, the dataset we collected is still insufficient and other existing published datasets, e.g., AMIGOS~\cite{shu2018review}, Augsburg Biosignal Toolbox (AuBT)~\cite{wagner2005augsburg}, etc., have not been verified by the proposed X-GWO-SVM algorithm. In future work, we will use other exploration-exploitation regulation functions for the proposed algorithm to explore
their effectiveness in emotion recognition. Additionally, more published datasets will be examined by our method.

Besides, through the results and conclusions reported in ~\cite{sarkar2020self}, we also observed that deep learning is competitive in emotion recognition, which may further improve the performance of our proposed strategy. Thus in our future work, we will try to find an effective deep learning method and embedded GWO methods to further improve emotion recognition performance.

\section{Conclusion}
\label{conclusion}
In this paper, we presented an X-GWO-SVM technique that improves the exploration and exploitation abilities of single channel ECG-based emotion recognition. In order to classify different emotions, this research used two reliable datasets: one trustworthy WESAD dataset and one raw self-collected iRealcare dataset. The single channel ECG signals could well be employed in the X-GWO-SVM algorithm for emotion recognition, according to 10-fold cross-validation results from 5 subjects for the iRealcare dataset and 15 subjects for the WESAD dataset. The algorithm performed better than past efforts that used various supervised machine learning techniques. It also provides a way to implement in the lightweight embedded system, which is much more efficient than existing solutions of using deep neural networks. The method has the potential to be used in clinical settings and also fills a gap in GWO-SVM research on ECG-based emotion identification. In our future work, we will apply radio sensing techniques, such as \cite{IREALCARE5,GI1,GI2,GI3,GI4}, instead of wearable devices for emotion recognition. We will also develop privacy preservation algorithms \cite{privacy1,privacy2,privacy3,privacy4} to protect the users' privacy.

%\bibliographystyle{IEEEtran}
%\bibliography{reference}

\end{document}